\documentclass[preprintnumbers,article,amsmath,amssymb,floatfix,10pt,prd,onecolumn,superscriptaddress,nofootinbib]{revtex4-2}

\usepackage{titlesec}
\usepackage[toc]{appendix}
\usepackage{caption}
\titleformat*{\section}{\LARGE\bfseries}
\titleformat*{\subsection}{\Large\bfseries}
\titleformat*{\subsubsection}{\large\bfseries}
\titleformat*{\paragraph}{\large\bfseries}
\titleformat*{\subparagraph}{\large\bfseries}
\usepackage{bm}
\usepackage{amsfonts}
\usepackage{latexsym}
\usepackage{graphicx}
\usepackage{amsmath}
\usepackage{palatino}
\usepackage{mathpazo}
\usepackage{textcomp}
\usepackage{float}
\usepackage{booktabs}
\usepackage{dcolumn}
\usepackage{ragged2e}
\usepackage{hyperref}
\hypersetup{colorlinks,citecolor=blue}
\hypersetup{colorlinks=true,linkcolor=blue,filecolor=magenta,    urlcolor=blue}
\usepackage{amsthm}
\usepackage{xcolor}
\usepackage{orcidlink}
\usepackage{epsfig}
\usepackage{subcaption}
\usepackage{commath}
\usepackage{cancel, ulem}
\usepackage[utf8]{inputenc}
\usepackage{multirow}
\def\jnl@style{\it}
\def\aaref@jnl#1{{\jnl@style#1}}

\def\aaref@jnl#1{{\jnl@style#1}}

\def\aj{\aaref@jnl{AJ}}                   
\def\apj{\aaref@jnl{ApJ}}                 
\def\apjl{\aaref@jnl{ApJ}}                
\def\apjs{\aaref@jnl{ApJS}}               
\def\apss{\aaref@jnl{Ap\&SS}}             
\def\aap{\aaref@jnl{A\&A}}                
\def\aapr{\aaref@jnl{A\&A~Rev.}}          
\def\aaps{\aaref@jnl{A\&AS}}              
\def\mnras{\aaref@jnl{Mon.~Not.~Roy.~Astron.~Soc.}}             
\def\prd{\aaref@jnl{Phys.~Rev.~D}}        
\def\prc{\aaref@jnl{Phys.~Rev.~C}}  
\def\prl{\aaref@jnl{Phys.~Rev.~Lett.}}    
\def\qjras{\aaref@jnl{QJRAS}}             
\def\skytel{\aaref@jnl{S\&T}}             
\def\ssr{\aaref@jnl{Space~Sci.~Rev.}}     
\def\zap{\aaref@jnl{ZAp}}                 
\def\nat{\aaref@jnl{Nature}}              
\def\aplett{\aaref@jnl{Astrophys.~Lett.}} 
\def\apspr{\aaref@jnl{Astrophys.~Space~Phys.~Res.}} 
\def\physrep{\aaref@jnl{Phys.~Rep.}}      
\def\physscr{\aaref@jnl{Phys.~Scr}}       
\def\commat{\aaref@jnl{Comm.~Math.~Phys.}}              
\def\science{\aaref@jnl{Science}}               
\def\cqg{\aaref@jnl{Classical Quant.~Grav.}}            
\def\jpcs{\aaref@jnl{JPCS}}                                     
\def\ijmpd{\aaref@jnl{Int.~J.~Mod.~Phys.~D}}                    
\def\grg{\aaref@jnl{Gen.~Relat.~Gravit.}}               
\def\rpp{\aaref@jnl{Rep.~Prog.~Phys.}}          
\def\npa{\aaref@jnl{Nucl.~Phys.~A}}        
\def\lrr{\aaref@jnl{Living Rev.~Rel.}}                   
\def\jcap{\aaref@jnl{J.~Cosmology Astropart.~Phys.}}    
\def\rmp{\aaref@jnl{Rev.~Mod.~Phys.}}   
\def\epjc{\aaref@jnl{Eur.~Phys.~J.~C}} 
\def\plb{\aaref@jnl{~Phy.~Lett.~B}} 
\def\mpla{\aaref@jnl{Mod.~Phy.~Lett.~A}} 
\def\arxiv{\aaref@jnl{arxiv.org}}

\allowdisplaybreaks[1]

\begin{document}

\title{The role of anisotropy in $f(Q)$ gravity: insights from cosmological observations}
\author{Ghulam Murtaza\orcidlink{0009-0002-6086-7346}}
\email{ghulammurtaza@1utar.my}
\affiliation{Department of Mathematical and Actuarial Sciences, Universiti Tunku Abdul Rahman, Jalan Sungai Long,
43000 Cheras, Malaysia}
\author{Nandan Roy\orcidlink{0000-0001-7197-453X}}
\email{nandan.roy@mahidol.ac.th}
\affiliation{NAS, Centre for Theoretical Physics \& Natural Philosophy, Mahidol University,
Nakhonsawan Campus, Phayuha Khiri, Nakhonsawan 60130, Thailand}
\author{Avik De\orcidlink{0000-0001-6475-3085}}
\email{avikde@um.edu.my}
\affiliation{Institute of Mathematical Sciences, Faculty of Science, Universiti Malaya, 50603 Kuala Lumpur, Malaysia}


\footnotetext{GM acknowledges Universiti Tunku Abdul Rahman Research Fund project IPSR/RMC/UTARRF/2023-C1/A09 provided by Universiti Tunku Abdul Rahman. AD acknowledges the Universiti Malaya BKP-ECRG Grant (Project No. BKP119-2025-ECRG).}

\begin{abstract}
We investigate the cosmological dynamics of Bianchi-I spacetime in symmetric teleparallel $f(Q)$ gravity through a dynamical system approach to analyse observational constraints. By reformulating the modified field equations into an autonomous system, we analyse two representative $f(Q)$ models and constrain their parameters using Pantheon Plus, DES Y5, DESI DR2, and compressed CMB data. The observational analysis yields consistent constraints across all dataset combinations and tightly bounds the anisotropic contribution, indicating that deviations from isotropy remain small. Both models reproduce the standard matter-dominated evolution and the observed late-time accelerated expansion while exhibiting distinct dark-energy dynamics. Model I undergoes a smooth phantom-divide crossing and approaches a de Sitter phase in the asymptotic future, whereas Model II evolves from an early phantom regime toward a cosmological-constant-like state around the present epoch, closely mimicking the late-time evolution of the $\Lambda$CDM model. These results indicate that anisotropic $f(Q)$ cosmology remains consistent with current background observations while admitting characteristic dark-energy evolution that may be testable with future cosmological surveys.
\end{abstract}

\maketitle
\tableofcontents
\section{Introduction}\label{sec00}
Unveiling the physical origin and dynamical nature of dark energy remains one of the foremost challenges in modern cosmology. The standard cosmological model, formulated within the homogeneous and isotropic Friedmann--Lemaître--Robertson--Walker (FLRW) spacetime, has achieved remarkable success in explaining a broad spectrum of observations. However, this framework fundamentally relies on the Cosmological Principle (CP), the postulate that the universe is statistically homogeneous and isotropic on sufficiently large scales \cite{Weinberg1998}. A growing body of observational evidence has motivated renewed scrutiny of this foundational assumption, particularly its assertion of statistical isotropy \cite{Schwarz2016,Bull2016,Perivolaropoulos2022,Abdalla2022}. While cosmic microwave background (CMB) measurements are broadly consistent with the standard model \cite{Bennett1996,Aghanim2020}, several large-angle anomalies and anisotropic features have been reported and extensively investigated \cite{Hinshaw2011,Ade2014,Ashdown2020,8,10,11,17}.

Beyond the CMB, tests of statistical isotropy have been conducted using diverse independent probes, including Type Ia supernovae \cite{Schwarz2007,Mohayaee2011,Wiltshire2013,Appleby2015,Zhou2019}, quasars \cite{Hutsemekers1998,Secrest2021}, radio polarization vector alignments \cite{Birch1982,Tiwari2013}, and large-scale peculiar velocity flows \cite{Watkins2009,Valade2024}. Intriguingly, several of these studies report preferred directions that exhibit notable alignment, approximately pointing towards the Virgo cluster \cite{Ralston2004}. Although none individually constitutes definitive evidence against the CP, their collective presence has stimulated considerable interest in cosmological models that permit small departures from exact isotropy. The unprecedented precision and abundance of modern cosmological data thus make testing statistical isotropy an increasingly important endeavor.

Within the standard spatially flat $\Lambda$CDM paradigm, the late-time accelerated expansion is attributed to a cosmological constant
($\Lambda$) with the equation of state (EoS) parameter 
$w_{\rm de}=-1$ \cite{Planck2018}. While this elegantly simple description provides an excellent fit to observations, it inherently assumes that the accelerated expansion proceeds identically in every spatial direction. Given the mounting evidence for possible departures from cosmic isotropy, it is natural to investigate whether the dark energy driving cosmic acceleration itself may exhibit a small degree of anisotropy. Addressing this question not only provides a direct test of the CP but also offers a valuable avenue to explore extensions of the standard model and probe the underlying physics governing late-time cosmic evolution.

These considerations motivate the exploration of cosmological models that relax exact isotropy while preserving spatial homogeneity. The simplest class of such models is provided by the Bianchi cosmologies \cite{MacCallum1969,Ellis1970}, which constitute homogeneous but anisotropic solutions of Einstein's field equations. Among these, the Bianchi type-I spacetime represents the simplest and most direct generalization of the spatially flat FLRW model. Owing to its spatial flatness and mathematical simplicity, it provides an ideal setting for examining the interplay between anisotropic geometry and matter-energy content, while retaining the essential features required to describe cosmic evolution. The line element of the Bianchi type-I spacetime is given by
\begin{equation}
ds^{2} = -dt^{2} + a_{1}^{2}(t)dx^{2} + a_{2}^{2}(t)dy^{2} + a_{3}^{2}(t)dz^{2},
\label{eq:bianchi1}
\end{equation}
where $a_{1}(t)$, $a_{2}(t)$, and $a_{3}(t)$ denote the directional scale factors along the $x$, $y$, and $z$ axes, respectively. Unlike the FLRW metric, which is characterized by a single universal scale factor, the Bianchi type-I spacetime admits independent expansion rates along three orthogonal spatial directions, enabling a quantitative description of anisotropic cosmic expansion.

Despite the remarkable success of General Relativity (GR), explaining the observed late-time acceleration necessitates the introduction of an exotic dark energy component whose physical origin remains elusive. Persistent challenges, including the Hubble tension \cite{Planck2018}, the 
$\sigma_8$ tension \cite{29}, and the cosmic coincidence problem \cite{32,33}, continue to motivate alternative descriptions of gravity. One promising direction is offered by modified theories of gravity, in which the geometric sector of Einstein's theory is generalized. Besides curvature-based extensions such as $f(\mathring{R})$ gravity \cite{34}, considerable attention has recently shifted towards teleparallel formulations, where gravitation is described by torsion or non-metricity instead of curvature. Among these, $f(Q)$ gravity has emerged as a particularly attractive framework. Based on symmetric teleparallel geometry, it attributes gravitation to the non-metricity scalar 
$Q$, defined through an affine connection with vanishing curvature and torsion, and generalizes the symmetric teleparallel equivalent of GR by replacing 
$Q$ with an arbitrary function $f(Q)$ \cite{surveyfQ}. Owing to its comparatively simple field equations and rich cosmological phenomenology, $f(Q)$ gravity has been extensively explored, yielding viable descriptions of both early and late-time universe evolution.

 For instance, in \cite{Basilakos2019}, the dynamical system analysis (DSA) of two viable $f(Q)$ models demonstrated the existence of a matter-dominated saddle point followed by a stable dark-energy-dominated attractor, while their growth history, quantified through $f\sigma_8$, remains consistent with observations and closely mimics the predictions of $\Lambda$CDM. Similarly, the exponential $f(Q)$ model investigated in \cite{Bohmer2023} exhibits a viable cosmic sequence, including radiation domination, a matter-dominated saddle era, and a stable de Sitter late-time attractor. In \cite{Paliathanasis2023}, the authors explored two families of symmetric flat connections in spatially flat FLRW $f(Q)$ gravity and showed that the de Sitter solution appears as a universal late-time attractor, while scaling solutions can successfully recover the standard radiation and matter-dominated epochs. From the observational perspective, several studies have demonstrated the capability of $f(Q)$ gravity to provide competitive alternatives to $\Lambda$CDM. In \cite{Lazkoz2019}, a redshift-based reconstruction of $f(Q)$ gravity using polynomial parameterizations and various cosmological observations revealed its potential to describe the late-time accelerated expansion without relying on the standard cosmological constant. Using EoS parametrizations and Markov Chain Monte Carlo (MCMC) analyses with different observational datasets, \cite{Koussour2023} showed that two $f(Q)$ models can reproduce viable quintessence-like evolution and provide fits comparable to $\Lambda$CDM, with their accelerating behavior further supported by the evolution of energy density, pressure, and the deceleration parameter. More recently, a theoretical and observational investigation of Hybrid $f(Q)$ gravity in \cite{Kolhatkar2026} demonstrated that early universe constraints fix the linear coupling, leading to a background expansion degenerate with $\Lambda$CDM, while modified perturbation dynamics generate distinctive signatures at late times. The inclusion of growth observations revealed a compensation mechanism affecting $\sigma_8$, with statistical analysis indicating a moderate-to-weak preference for the Hybrid scenario and providing possible signatures for future large-scale structure observations. Several other important developments in this direction can be found in \cite{DAmbrosio2022,Giacomini2026,De2026,52,56,60}.

Nevertheless, most of these investigations are restricted to isotropic cosmological backgrounds, whereas anisotropic cosmologies have received comparatively limited attention. In particular, Bianchi type-I cosmology in $f(Q)$ gravity has been recently explored. The first study of anisotropic cosmology in $f(Q)$ gravity, based on an LRS Bianchi-I spacetime, was presented in \cite{firstBI} to investigate the early evolution of the universe. The study showed that significant anisotropy may be present in the early universe, while the cosmic evolution can progressively approach isotropy, with the anisotropic effects of the fluid also diminishing over time. In \cite{2}, anisotropic cosmological models in $f(Q)$ gravity were investigated, suggesting that anisotropy may play a significant role in the early evolution of the universe, while its effects become negligible at the present epoch. The strength and evolution of these anisotropic effects depend on the choice of the $f(Q)$ model and its associated model parameters. Similarly, \cite{Rathore:2024ema} analyzed an LRS Bianchi-I quadratic $f(Q)$ model and found unstable quintessence and phantom-like dark-energy solutions together with an attractor solution describing the asymptotic evolution. In \cite{Leon2024}, a minisuperspace formulation followed by a dynamical system approach was employed to study Bianchi-I cosmology in $f(Q)$ gravity, with particular emphasis on Kasner and Kasner-like solutions and the influence of different connections, especially in the regimes where the non-metricity scalar vanishes. Furthermore, using a covariant formulation combined with DSA, \cite{Esposito:2022omp} investigated Bianchi-I solutions and demonstrated how different classes of $f(Q)$ models can recover isotropic cosmological epochs. In our previous work \cite{ghulam2025}, we developed a general dynamical system framework for Bianchi type-I cosmology in $f(Q)$ gravity and demonstrated that a broad class of models naturally admits a physically viable de Sitter late-time attractor. However, the aforementioned studies, including our previous analysis, mainly focused on the theoretical viability of the models through the structure of the phase space, critical points, and their stability properties. Consequently, the allowed parameter space was primarily determined from dynamical considerations, without a direct confrontation with current cosmological observations.

Motivated by this, the present work extends the previous dynamical analysis by performing a comprehensive observational investigation of the considered $f(Q)$ models using combinations of recent cosmological datasets. Our aim is to examine whether the theoretically predicted cosmic evolution is supported by observations, constrain the model parameters within observationally viable regions, and establish a consistent picture by combining dynamical-system predictions with cosmological data analysis.

Although dynamical system analysis is traditionally employed to investigate the qualitative phase-space structure and asymptotic behavior of cosmological models, its formulation in terms of normalized dimensionless variables also provides a powerful framework for observational studies. By expressing the cosmological equations as a closed autonomous system, the evolution of the universe can be computed in a numerically stable and computationally efficient manner over a wide range of redshifts. In this work, we exploit this formulation not for a fixed-point analysis, but as the theoretical basis for Bayesian parameter estimation. Specifically, the autonomous system is solved numerically and embedded within an MCMC pipeline to constrain the free parameters using cosmological observations. This methodology allows the entire cosmic evolution to be reconstructed consistently for each sampled parameter set, enabling direct comparisons with observational data while simultaneously providing predictions for physically relevant quantities. The combination of the dynamical system framework with Bayesian statistical inference thus offers a robust and self-consistent approach for testing the observational viability of cosmological models \cite{Hussain2024,Paliath2026,Niyogi2024,Cedeno2021}.

Building upon these foundations, in this work, we consider the Bianchi type-I spacetime in the framework of symmetric teleparallel $f(Q)$ gravity and examine two specific models, confronting them with a broad range of observational datasets. These include CMB data, Baryon Acoustic Oscillation (BAO) measurements from the Dark Energy Spectroscopic Instrument (DESI) Release II, and different Type Ia Supernova samples, namely Pantheon+ and the Dark Energy Survey five-year (DESY5) compilation. The paper is organized as follows. After this introduction, Section~\ref{sec1} provides a brief mathematical framework for $f(Q)$ theory. In Section~\ref{seciv} we rewrite the autonomous dynamical system equations for Bianchi-I cosmology in $f(Q)$ framework, and in the subsections, introduce the two particular models under consideration, and outline the observational datasets to constrain them. Our results and their physical implications are presented and discussed in Section~\ref{V}. Finally, Section~\ref{VI} summarizes our main findings and concluding remarks.

\section{Fundamentals of $f(Q)$ Gravity}\label{sec1}
In symmetric teleparallel gravity, gravitation is described by the non-metricity of spacetime rather than curvature or torsion. Accordingly, the affine connection satisfies
\begin{equation}
R^{\rho}{}_{\sigma\mu\nu}=0,\qquad
T^{\rho}{}_{\mu\nu}=0,\qquad
Q_{\lambda\mu\nu}\equiv\nabla_{\lambda}g_{\mu\nu}\neq0,
\end{equation}
where $Q_{\lambda\mu\nu}$ denotes the non-metricity tensor.

The affine connection can be decomposed into the Levi--Civita connection and the disformation tensor as
\begin{equation}
\Gamma^{\lambda}{}_{\mu\nu}
=\mathring{\Gamma}^{\lambda}{}_{\mu\nu}
+L^{\lambda}{}_{\mu\nu},
\end{equation}
with
\begin{equation}
L^{\lambda}{}_{\mu\nu}
=\frac12\left(
Q^{\lambda}{}_{\mu\nu}
-Q_{\mu}{}^{\lambda}{}_{\nu}
-Q_{\nu}{}^{\lambda}{}_{\mu}
\right).
\end{equation}

The non-metricity scalar is constructed from the superpotential
\begin{equation}
P^{\lambda}{}_{\mu\nu}
=\frac14\left(
-2L^{\lambda}{}_{\mu\nu}
+Q^{\lambda}g_{\mu\nu}
-\tilde Q^{\lambda}g_{\mu\nu}
-\delta^{\lambda}_{(\mu}Q_{\nu)}
\right),
\end{equation}
through
\begin{equation}
Q=Q_{\lambda\mu\nu}P^{\lambda\mu\nu},
\end{equation}
where
\begin{equation}
Q_{\lambda}=Q_{\lambda\mu}{}^{\mu},
\qquad
\tilde Q_{\lambda}=Q^{\mu}{}_{\lambda\mu}.
\end{equation}

The action for $f(Q)$ gravity is given by
\begin{equation}
S=\int
\left[
\frac{f(Q)}{2\kappa}
+\mathcal{L}_{m}
\right]
\sqrt{-g}\,d^{4}x,
\end{equation}
where $\kappa=8\pi G$ and $\mathcal{L}_{m}$ denotes the matter Lagrangian.

Variation of the action with respect to the metric yields the field equations
\begin{equation}
f_Q\mathring{G}_{\mu\nu}
+\frac12g_{\mu\nu}
\left(
Qf_Q-f
\right)
+2f_{QQ}
\mathring{\nabla}_{\lambda}Q
P^{\lambda}{}_{\mu\nu}
=\kappa T_{\mu\nu},
\label{FE2}
\end{equation}
where $\mathring{G}_{\mu\nu}$ is the Einstein tensor constructed from the Levi--Civita connection, $f_Q= df/dQ$ and $f_{QQ}=d^{2}f/dQ^{2}$.
Since the affine connection is treated as an independent geometrical quantity in symmetric teleparallel gravity, variation of the action with respect to the connection $\Gamma^{\lambda}_{\mu\nu}$ leads to an additional field equation given by
\begin{equation}
\nabla_{\mu}\nabla_{\nu}\left(\sqrt{-g}f_Q P^{\mu\nu}_{\ \ \gamma}\right)=0.
\end{equation}
This equation is obtained under the assumption that the matter Lagrangian $\mathcal{L}_{m}$ does not explicitly depend on the affine connection. It is worth noting that the construction of $f(Q)$ gravity is based on the restrictions of vanishing curvature and torsion, namely $R^{\rho}_{\ \sigma\mu\nu}=0$ and $T^{\rho}_{\ \mu\nu}=0$. These conditions imply the existence of a coordinate system in which the affine connection coefficients can be set to zero, $\Gamma^{\lambda}_{\mu\nu}=0$, which is known as the coincident gauge. This choice considerably simplifies the geometrical description of symmetric teleparallel gravity while retaining its full dynamical content. Notably, it has been demonstrated in \cite{Loo/2023} that for the class of spacetime geometries considered in this work, the connection field equation is identically satisfied irrespective of the specific form of the function $f(Q)$. Therefore, throughout this study, we focus solely on the metric field equations provided in Section \ref{seciv}.


\section{Autonomous System of Bianchi-I Cosmology in $f(Q)$ Gravity}\label{seciv}

Following our previous work \cite{ghulam2025}, we employ the dynamical system formulation for a spatially homogeneous Bianchi-I universe in $f(Q)$ gravity. Since the complete derivation has already been presented in detail in \cite{ghulam2025}, we only summarize the essential ingredients required for the subsequent analysis.

The Bianchi-I spacetime is described by the metric (\ref{eq:bianchi1}), with directional scale factors parameterized as
\begin{equation}
a_i=a\,e^{\beta_i}, \qquad
\sum_{i=1}^{3}\beta_i=0,
\end{equation}
where $a=(a_1a_2a_3)^{1/3}$ is the average scale factor. The directional Hubble parameters are therefore
\begin{equation}
H_i=H+\dot{\beta}_i,
\end{equation}
with $H=\dot a/a$ denoting the mean Hubble parameter. The anisotropy is characterized by the shear scalar
\begin{equation}
\sigma^2=\sum_{i=1}^{3}\dot{\beta}_i^{\,2},
\end{equation}
which vanishes in the isotropic FLRW limit.

Throughout this work, we assume an isotropic perfect fluid,
\begin{equation}
T^\nu_{\ \mu}=\mathrm{diag}(-\rho,P,P,P), \qquad P=\omega\rho,
\end{equation}
while allowing the spacetime geometry to possess a small shear contribution. This approximation is well motivated for late-time cosmology and has been discussed in detail in \cite{Chakraborty2022}.

The Bianchi-I cosmological field equations in the presence of an isotropic fluid in the context of $f(Q)$ gravity ($Q=\sigma^2-6H^2$) are given by
\begin{align}\label{1eq:m}
3H^2 - \frac{\sigma^2}{2} = \frac{\kappa}{2f_Q}\left(\rho - \frac{f}{2\kappa}\right), 
\end{align}
\begin{align}\label{2eq:m}   
-(\dot H+3H^2) = \frac{\kappa}{2f_Q}\left(\omega \rho + \frac{f}{2\kappa}+\frac{2H\dot{f_Q}}{\kappa}\right)\, .
\end{align}
supplemented by the shear evolution equation
\begin{align}\label{connllsigma}
     \dot{\sigma}= -\sigma\left(\frac{\dot{f}_{Q}}{f_{Q}}+3H\right).
\end{align}
The standard conservation equation is retained.
\begin{align}
    \dot \rho +3H(\rho+P)=0.
\end{align}
Introducing the dimensionless variables
\begin{align}
   x_{1} = \frac{\kappa \rho}{6f_{Q}H^{2}}, \qquad x_{2} = -\frac{f}{12f_{Q}H^{2}}, \qquad x_{3} = \frac{\sigma^{2}}{6H^{2}},
   \label{variables}
\end{align}
the equation (\ref{1eq:m}) immediately yields the constraint
\begin{align} \label{constraint}
    1=x_{1}+x_{2}+x_{3}.
\end{align}
one can also write
\begin{align}
    \Omega_{DE}=1+2f_Q(-1+x_3+x_2) , ~~\Omega_m=2x_1f_Q.
\end{align}

The autonomous dynamical system equations can be obtained as,
\begin{align}
    x'_{1} = -3x_{1}(1+\omega)+6x_{1}\left(\frac{(x_{2}-1-x_{1}\omega)(1+\Gamma-\Gamma x_{3}-2x_{3})}{(\Gamma +2)(x_{3}-1)}\right)-\frac{6x_{1}x_{3}}{(\Gamma +2)(x_{3}-1)},
\end{align} 
\begin{align}
    x'_{2} = 3\Gamma \left(\frac{x_{3}+x_{2}-1-x_{1}\omega}{\Gamma +2}\right)+6x_{2}\left(\frac{(x_{2}-1-x_{1}\omega)(1+\Gamma-\Gamma x_{3}-2x_{3})}{(\Gamma +2)(x_{3}-1)}\right)-\frac{6x_{2}x_{3}}{(\Gamma +2)(x_{3}-1)},
\end{align}  
\begin{align}
  x'_{3} = 6x_{3}(x_{1}\omega -x_{2}).
\end{align}
In the above set of equations, prime $(.)'$ represents the derivative with respect to $N=\ln a$, and we define the auxiliary quantity $\Gamma=\Gamma(Q)$ as,
\begin{align}\label{Gamma}
\Gamma = \frac{f_{Q}}{Q f_{QQ}}.
\end{align}
Using the (\ref{2eq:m}), we have,
\begin{align}\label{cosmology}
\frac{\dot{H}}{H^{2}} = 3\left( \frac{(2x_{3}+x_{3}\Gamma -\Gamma)(x_{2}-1-x_{1}\omega )+2x_{3}}{(\Gamma+2)(-1+x_{3})}\right).
\end{align}

Using the constraint to eliminate the $x_{2}$, dynamical system equations reduce to the following way,
\begin{align}\label{gds1}
x'_{1} = -3x_{1}(1+\omega)+6x_{1}(x_{1}+x_{3}+x_{1}\omega)+6x_{1}\frac{x_{1}+x_{1}\omega}{(\Gamma +2)(x_{3}-1)},
\end{align}
\begin{align}\label{gds2}   
x'_{3} = 6x_{3}x_{1}(1+\omega)+6x^{2}_{3}-6x_{3},
\end{align}

while \eqref{cosmology} reduces to
\begin{align}\label{cosmology_new}
\frac{\dot{H}}{H^{2}} =  3\left( \frac{(2x_{3}+x_{3}\Gamma -\Gamma)(-x_{3}-x_{1}-x_{1}\omega )+2x_{3}}{(\Gamma+2)(-1+x_{3})}\right).
\end{align}
Using (\ref{variables}) and then using the constraint equation, we can express $Q$ as,
 \begin{align}\label{Qvar1}
    \frac{Q f_Q}{f} =  \frac{(1-x_3)}{2(1-x_1-x_3)}.
\end{align}
It is important to note that, by definition, the dimensionless variables $x_1$ and $x_3$ satisfy the non-negative condition $x_1 \geq 0$ and $x_3 \geq 0$. Therefore, the physical region is confined to this domain, and these constraints are consistently enforced throughout our analysis. In addition, we consider the matter component to be pressureless dust $(P = 0)$, which provides an appropriate description of the late-time matter sector of the universe.

\subsection{Model-I: $f(Q) = Q e^ {\lambda \frac{Q_{0}}{Q}}$}
The power-exponential $f(Q)$ model has been investigated in \cite{Anagnostopoulo2021,Anagnostop2023} and shown to satisfy the constraints imposed by the early universe, making it a viable candidate for describing cosmic evolution. Furthermore, its observational viability has been examined using a variety of cosmological datasets, demonstrating that its performance is comparable to the standard $\Lambda$CDM model for certain datasets, while for others it provides an even better statistical fit. The model contains a free parameter $\lambda$, which quantifies deviations from GR. In the limit $\lambda \to 0$, the theory reduces to GR without a cosmological constant. Although the model does not admit an exact $\Lambda$CDM limit, for sufficiently small values of $\lambda$ it effectively reproduces GR supplemented by an emergent cosmological constant of order $Q_0 \lambda$, thereby closely mimicking the late-time expansion history predicted by the $\Lambda$CDM paradigm. The forms of $f_Q$, $Q$, and $\Gamma(Q)$ in terms of variables for this example can be obtained as
\begin{align}
    f_Q=e^{\frac{\lambda Q_0}{Q}}\left(1-\frac{\lambda Q_0}{Q}\right), ~~ f_{QQ}=e^{\frac{\lambda Q_0}{Q}}\left(\frac{\lambda^2 Q_0^2}{Q^3}\right).
    \label{fq3}
\end{align}
Explicit substitution into (\ref{Qvar1}) gives
\begin{align}
    Q = \frac{2x_{2} \lambda Q_{0}}{2x_{2}-1+x_{3}},
    \label{Qe}
\end{align}
substituting  (\ref{Qe}) into (\ref{fq3}), one can get
\begin{align}
f_Q = e^{\frac{1 - x_3 - 2 x_1}{2 - 2 x_1 - 2 x_3}} \left( \frac{1 - x_3}{2 - 2 x_1 - 2 x_3} \right),
\end{align}
using (\ref{fq3}) and (\ref{Qe}) into (\ref{Gamma}), one can get
\begin{align}
   \Gamma = \frac{4(1-x_{1}-x_{3})^{2}}{(1-2x_{1}-x_{3})^{2}}-\frac{2 (1-x_{1}-x_{3})}{1-2x_{1}-x_{3}}.
\end{align}
Replacing this value of $\Gamma$ in (\ref{gds1}) and (\ref{cosmology_new}), and convert all derivatives with respect to $N$ to redshift $z$ using the relations
 $\frac{dx}{dz}=-\frac{x'}{1+z},~~ \frac{dH}{dz}=\frac{\dot{H}}{-(1+z)H}$, also assuming the pressureless dust era, we get the following,
 \begin{align} 
\frac{dx_1}{dz} = -\frac{1}{1+z}\left(-3x_1 + 6x_1(x_1 + x_3)
 + \frac{3x_1^2(1 - 2x_1 - x_3)^2}
{(x_3 - 1)\left[2(1 - x_1 - x_3)^2 - (1 - x_1 - x_3)(1 - 2x_1 - x_3) + (1 - 2x_1 - x_3)^2\right]}\right),
\end{align}
\begin{align}
\frac{dx_3}{dz} =-\frac{1}{1+z}\left( 6x_3x_1+6x^{2}_{3}-6x_{3}\right). 
\end{align}
It is evident that the dynamical system equations of this model are independent of the model parameter.
{\scriptsize
\begin{align}
    \frac{dH}{dz}=-\frac{3H}{1+z}\left(\frac{\left(2x_3(1-2x_1-x_3)^2 +\left(4(1-x_1-x_3)^2-2(1-x_1-x_3)(1-2x_1-x_3)\right)(x_3-1)\right)(-x_3-x_1)+2x_3(1-2x_1-x_3)^2}{\left(4(1-x_1-x_3)^2-2(1-x_1-x_3)(1-2x_1-x_3)+2(1-2x_1-x_3)^2\right)(x_3-1)}\right).
\end{align}
}

We can express the effective equation of state, $w_{eff}=-1-\frac{2 \dot H}{3H^2}$ as,
{\scriptsize
\begin{align}
   w_{eff}=-1-2\left(\frac{\left(2x_3(1-2x_1-x_3)^2 +\left(4(1-x_1-x_3)^2-2(1-x_1-x_3)(1-2x_1-x_3)\right)(x_3-1)\right)(-x_3-x_1)+2x_3(1-2x_1-x_3)^2}{\left(4(1-x_1-x_3)^2-2(1-x_1-x_3)(1-2x_1-x_3)+2(1-2x_1-x_3)^2\right)(x_3-1)}\right).   
\end{align}
}
and also the dark energy equation of state, $w_{DE} \Omega_{DE}=-1-\frac{2 \dot H}{3H^2}$ as,
{\scriptsize
\begin{align}
w_{DE} 
&= \frac{x_3+x_1-1}{1-x_1-x_3-x_1e^{\frac{1-x_3-2x_1}{2-2x_1-2x_3}}(1-x_3)} - \frac{2(1-x_3-x_1)}{1-x_1-x_3-x_1 e^{\frac{1-x_3-2x_1}{2-2x_1-2x_3}}(1-x_3)}\\\nonumber
&
\quad\times \Bigg( 
\frac{
\left(2x_3(1-2x_1-x_3)^2 
+\left(4(1-x_1-x_3)^2 -2(1-x_1-x_3)(1-2x_1-x_3)\right)(x_3-1)\right)
(-x_3-x_1)
+2x_3(1-2x_1-x_3)^2
}{
\left(4(1-x_1-x_3)^2 
-2(1-x_1-x_3)(1-2x_1-x_3)
+2(1-2x_1-x_3)^2\right)(x_3-1)
}
\Bigg).
\end{align}
}
\subsection{Model-II: $f(Q) = Q + \alpha \sqrt {-Q} +\Lambda$}\label{model3}
This particular $f(Q)$ model has been investigated previously in \cite{m2,Chakraborty:2025qlv}, and it yields a real-valued gravitational Lagrangian $f(Q)$ capable of exactly reproducing the $\Lambda$CDM expansion history for a universe dominated by pressureless (dust) matter. An appealing feature of this model is that, in the limit $\Lambda \to 0$, it naturally reduces to a dust-dominated cosmological evolution while remaining distinct from GR, thereby providing a genuinely modified gravity realization of the matter-dominated era.
We can write the expressions for $f_Q$, $Q$, and $\Gamma(Q)$ in terms of the dynamical variables in the following way,
\begin{align}
    f_Q=1-\frac{\alpha}{2\sqrt{-Q}}, ~~ f_{QQ}=-\frac{\alpha}{4(-Q)^\frac{3}{2}}.
    \label{fq2}
\end{align}
Explicit substitution into (\ref{Qvar1}) gives
\begin{align}
    \sqrt{-Q} = \frac{\alpha x_{1} \pm \sqrt{\alpha^{2} x_{1}^{2}-4(x_{3}-1+2x_{1})(\Lambda x_{3}-\Lambda)}}{2(x_{3}+2x_{1}-1)}.
    \label{negQ}
\end{align}
We restrict our analysis to the positive branch of the above expression. Putting (\ref{negQ}) into (\ref{fq2}), one can get
\begin{align}
f_{Q}=\frac{(1-x_{3}-x_{1})+ \sqrt{x_{1}^{2}-4\frac{\Lambda}{\alpha^2}(x_{3}+2x_{1}-1)(x_{3}-1)}}{x_{1} + \sqrt{x_{1}^{2}-4\frac{\Lambda}{\alpha^2}(x_{3}+2x_{1}-1)(x_{3}-1)}}, 
\end{align}
Utilizing (\ref{fq2}) and (\ref{negQ}) in (\ref{Gamma}), we have

\begin{align}
   \Gamma = \frac{2 x_{1} + 2\sqrt{x_{1}^{2}-4\frac{\Lambda}{\alpha^2}(x_{3}-1+2x_{1})( x_{3}-1)}}{ (x_{3}+2x_{1}-1)} -2 .
\end{align}
Since $x_1$ and $x_3$ are non-negative and for $\Gamma$ to be real-valued, we obtain the following constraint on the model parameters, i.e., $\alpha \neq 0$ and $\Lambda \leq 0$. Substituting this expression for $\Gamma$ into (\ref{gds1}) and (\ref{cosmology_new}), transforming the derivatives from the e-folding variable $N$ to the redshift $z$, and assuming a pressureless dust-dominated universe, we obtain the following equations

\begin{align} 
\frac{dx_1}{dz} = -\frac{1}{1+z}\left(-3x_{1} +6x_{1}(x_{1}+x_{3})+ \frac{6x_{1}^2 (x_{3}+2x_{1}-1)}{(x_{3}-1)\left(2x_{1} + 2 \sqrt{x_{1}^{2}-4\frac{\Lambda}{\alpha^2}(x_{3}-1+2x_{1})(x_{3}-1)}\right)}\right),
\end{align}
\begin{align}
\frac{dx_3}{dz} =-\frac{1}{1+z}\left( 6x_3x_1+6x^{2}_{3}-6x_{3}\right). 
\end{align}

{\scriptsize
\begin{align}
   \frac{dH}{dz}=-\frac{3H}{1+z}\left(\frac{\left(2x_3(x_3+2x_1-1)+\left(2x_1+2\sqrt{x_1^2-4\frac{\Lambda}{\alpha^2}(x_3-1+2x_1)(x_3-1)}-2(x_3+2x_1-1)(x_3-1)\right)\right)(-x_3-x_1)+2x_3(x_3+2x_1-1)}{\left(2x_1+2\sqrt{x_1^2-4\frac{\Lambda}{\alpha^2}(x_3-1+2x_1)(x_3-1)}\right)(x_3-1)}\right).
\end{align}
}

The expression for $w_{eff}$ can be written as,
{\scriptsize
\begin{align}
   w_{eff}=-1-2\left(\frac{\left(2x_3(x_3+2x_1-1)+\left(2x_1+2\sqrt{x_1^2-4\frac{\Lambda}{\alpha^2}(x_3-1+2x_1)(x_3-1)}-2(x_3+2x_1-1)(x_3-1)\right)\right)(-x_3-x_1)+2x_3(x_3+2x_1-1)}{\left(2x_1+2\sqrt{x_1^2-4\frac{\Lambda}{\alpha^2}(x_3-1+2x_1)(x_3-1)}\right)(x_3-1)}\right).
\end{align}
}
and for $w_{DE}$, one can get the following,
{\scriptsize
\begin{align}
w_{DE} 
&= -\frac{ x_1+\sqrt{ x_1^2-4\frac{\Lambda}{\alpha^2}(x_3-1+2x_1)(x_3-1)}}{\sqrt{ x_1^2-4\frac{\Lambda}{\alpha^2}(x_3-1+2x_1)( x_3-1)}+ x_1-2x_1\left(\sqrt{ x_1^2-4\frac{\Lambda}{\alpha^2}(x_3-1+2x_1)(x_3-1)}+(1-x_1-x_3)\right)} \\\nonumber
&\quad -2\left(\frac{\left(2x_3(x_3+2x_1-1)+\left(2x_1+2\sqrt{x_1^2-4\frac{\Lambda}{\alpha^2}(x_3-1+2x_1)(x_3-1)}-2(x_3+2x_1-1)(x_3-1)\right)\right)(-x_3-x_1)+2x_3(x_3+2x_1-1)}{\left(2x_1+2\sqrt{x_1^2-4\frac{\Lambda}{\alpha^2}(x_3-1+2x_1)(x_3-1)}\right)(x_3-1)}\right)\times\\\nonumber 
&\quad\frac{ x_1+\sqrt{ x_1^2-4\frac{\Lambda}{\alpha^2}(x_3-1+2x_1)( x_3-1)}}{\sqrt{x_1^2-4\frac{\Lambda}{\alpha^2}(x_3-1+2x_1)( x_3-1)}+ x_1-2x_1\left(\sqrt{ x_1^2-4\frac{\Lambda}{\alpha^2}(x_3-1+2x_1)( x_3-1)}+(1-x_1-x_3)\right)}.
\end{align}}

\subsection{Observational Data}\label{Observational Data}
To assess the observational viability of the proposed model, we perform a statistical analysis using a combination of current cosmological datasets. The model parameters are constrained through an MCMC analysis implemented within the \textit{COBAYA} framework~\cite{Torrado:2020dgo}. The background evolution equations are implemented within the \textit{CosmoDS} \cite{cosmods}, which serves as an external theory component for \textit{COBAYA} and is specifically designed for constraining cosmological models formulated through dynamical system analysis. The resulting posterior distributions are subsequently analyzed and visualized using the publicly available \textit{GetDist} package.


\subsubsection{Supernova Data}
Type Ia supernovae (SN-Ia) serve as reliable standard candles due to their nearly uniform intrinsic luminosities, making them one of the most powerful probes of the cosmic expansion history~\cite{reiss1998supernova,SupernovaSearchTeam:1998fmf}. In the present analysis, we employ the Pantheon Plus compilation~\cite{Scolnic:2021amr, Riess_2022, Brout:2022vxf} together with the DES Year 5 supernova sample~\cite{DES:2024jxu}. These independent compilations are based on distinct survey strategies, calibration procedures, and photometric analyses, collectively providing precise measurements of the distance modulus, $\mu$, across a wide range of redshifts.


\subsubsection{DESI BAO Data}
Baryon acoustic oscillations (BAO), imprinted by sound waves in the primordial baryon photon plasma, provide a robust standard ruler for measuring the expansion history of the universe. In the present work, we employ the latest BAO measurements from the 2025 Dark Energy Spectroscopic Instrument (DESI DR2) release~\cite{DESI:2025zgx, DESI:2024mwx}\footnote{The DESI DR2 dataset used here is available at \href{https://github.com/CobayaSampler}{https://github.com/CobayaSampler}.}.

BAO observations constrain distances both along and across the line of sight. The radial (parallel) distance measure is given by
\begin{equation}
\frac{D_H(z)}{r_d} = \frac{c}{H(z) r_d} \ ,
\end{equation}
while the transverse (perpendicular) comoving distance is expressed as,
\begin{equation}
\frac{D_{M}(z)}{r_{d}} = \frac{c}{r_{d}} \int_{0}^{z} \frac{d\tilde{z}}{H(\tilde{z})} = \frac{c}{H_{0} r_{d}} \int_{0}^{z} \frac{d\tilde{z}}{h(\tilde{z})}.
\end{equation}
An additional isotropic distance estimator, obtained by averaging over angles, is defined as,
\begin{equation}
\frac{D_V(z)}{r_d} = \left[\frac{c z ~ d_L^2(z)}{H(z) ~ r_d^{3} (1+z)^2}\right]^{\frac{1}{3}} \ ,
\end{equation}
where $d_L(z)$ denotes the luminosity distance. In this work, we consider $r_d$ as a free parameter.

\subsubsection{Compressed CMB Likelihood}
In addition to late-time cosmological observations, we incorporate constraints from CMB through the compressed likelihood formalism, which efficiently captures the background cosmological information most relevant for late-time parameter estimation. Following the procedure outlined in Appendix~A of ~\cite{DESI:2025zgx},
parameters are tightly constrained by CMB observations and can be determined with only weak sensitivity to late-time cosmological evolution after marginalizing over contributions such as gravitational lensing and the late integrated Sachs-Wolfe effect~\cite{Lemos:2023xhs}.

This compressed likelihood provides a robust high-redshift anchor for the cosmological expansion history while retaining nearly all of the information relevant to dark energy parameter estimation. In particular, it imposes stringent constraints on the matter density, thereby reducing parameter degeneracies that commonly arise when low-redshift probes, such as BAO, are considered. As demonstrated in ~\cite{DESI:2025zgx}, the cosmological constraints obtained with the compressed likelihood closely reproduce those derived from the full CMB likelihood, confirming that the dominant cosmological information is effectively preserved. Consequently, we adopt this compressed CMB likelihood as a computationally efficient and reliable alternative to a full CMB power-spectrum analysis.
We consider the following combinations of datasets:

\begin{enumerate}
\item Set 1: Pantheon Plus + CMB + DESI DR2,
\item Set 2: DES Y5 + CMB + DESI DR2,
\item Set 3: CMB + DESI DR2.
\end{enumerate}

\section{Physical Significance}\label{V}
{$\bullet$ \bf Model I:}  We constrain the model using three independent combinations of cosmological observations: (i) Pantheon Plus + CMB + DESI DR2, (ii) DES Y5 + CMB + DESI DR2, and (iii) CMB + DESI DR2. The free parameters varied in the MCMC analysis are $({H_0, x_{10}, x_{30}, r_d})$, with the adopted prior ranges listed in Table~\ref{tab:prior_range}. Since the dynamical variables $x_1$ and $x_3$ are required to satisfy the physical conditions $(x_1,x_3\geq0)$, the corresponding prior ranges are chosen accordingly. Furthermore, the parameter space is explored only within the region that yields physically viable cosmological solutions. The resulting marginalized posterior distributions are displayed in Fig.~\ref{fig1}, while the corresponding mean values and 68 \% confidence intervals are summarized in Table~\ref{tab:best_fit_values}.

The observational constraints obtained from the three dataset combinations are consistent with each other. For the Pantheon Plus + CMB + DESI DR2 combination, the model predicts $H_0\simeq71.74$ together with a present matter density parameter $\Omega_{m}\simeq0.28$. Using DES Y5 + CMB + DESI DR2 yields $H_0\simeq70.96$ and $\Omega_{m}\simeq0.29$, whereas the CMB + DESI DR2 combination favors $H_0\simeq71.93$ and $\Omega_{m}\simeq0.27$. In all three cases, the anisotropic parameter $x_{30}$ is tightly constrained to be of the order of $10^{-4}$, indicating that the present universe remains highly isotropic while allowing for a small but non-vanishing anisotropic contribution. Likewise, the sound horizon is consistently determined to be $r_d\simeq147.2$, exhibiting negligible variation among the three observational combinations and remaining fully consistent with current cosmological constraints.

\begin{figure}[t]
	\includegraphics[scale=0.5]{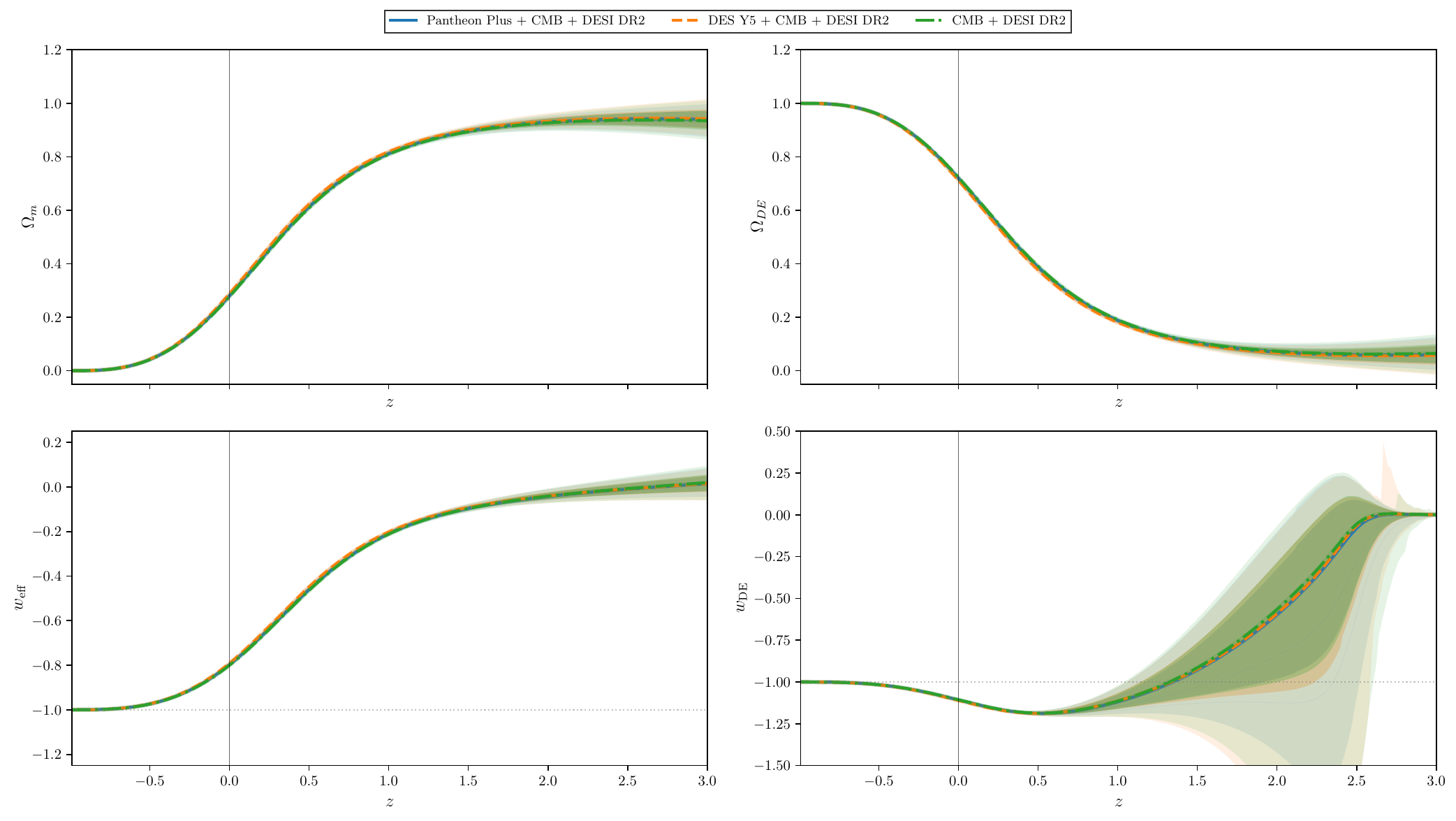}
	\caption{Evolution of the matter density parameter $\Omega_m$, dark energy density parameter $\Omega_{DE}$, effective equation of state parameter $w_{eff}$, and dark energy equation of state parameter $w_{DE}$ for Model I. The blue, orange, and green curves correspond to the Pantheon Plus+CMB+DESI DR2, DES Y5+CMB+DESI DR2, and CMB+DESI DR2 datasets, respectively. The associated dark and light shaded regions indicate the 68\% $(1\sigma)$ and 95\% $(2\sigma)$ confidence intervals.}
	\label{fig:params1}
\end{figure}

An equally notable feature is the stability of the derived cosmological parameters. The present effective EoS parameter is found to be $w_{ eff}\simeq-0.79$ for all dataset combinations, indicating a universe undergoing accelerated expansion. Similarly, the dark-energy EoS parameter is consistently constrained to $w_{ DE}\simeq-1.10$, suggesting a mild phantom behavior at the present epoch. These constraints are illustrated in Fig.~\ref{fig1}, while their redshift evolution is presented in Fig.~\ref{fig:params1}. 

The evolution of $w_{\rm DE}$ reveals particularly interesting dynamics. At high redshift, $w_{ DE}$ approaches zero, implying that the dark-energy component contributes negligibly during the matter-dominated era, thereby preserving the standard cosmological evolution. As the universe evolves, $w_{ DE}$ decreases and undergoes a phantom-divide crossing $w_{ DE}=-1$ around $z\approx1.4$. At the present epoch $z=0$, it reaches $w_{\rm DE}\simeq-1.10$, indicating a phantom dark-energy phase. In the asymptotic future, however, $w_{ DE}$ gradually approaches $-1$, demonstrating that the model naturally evolves toward a cosmological constant-like state. Consequently, the model predicts a smooth transition from a quintessence-like regime $w_{ DE}>-1$ to a phantom phase $w_{DE}<-1$, followed by a relaxation toward the $\Lambda$CDM limit at near future.

The remaining cosmological parameters exhibit similarly well-behaved evolution, i.e., $w_{eff}$ evolves toward $-1$, while the matter density parameter decreases monotonically from its matter-dominated value to the observed present value $\Omega_{m}\approx0.28$ before becoming negligible in the near future. Simultaneously, the dark-energy density parameter evolves monotonically from a negligible value at early times to the observed value $\Omega_{DE}\approx0.72$, before asymptotically approaching unity in the future, reflecting the eventual dark-energy-dominated phase of the universe. Overall, the cosmological evolution predicted by the model remains fully consistent with the standard thermal and matter-dominated history of the universe while introducing a distinctive late-time modification driven by the anisotropic $f(Q)$ sector; however, this phantom phase is transient rather than permanent. As the universe continues to expand, $w_{ DE}$ gradually approaches $-1$, and the cosmic evolution asymptotically converges to a de Sitter state that is observationally indistinguishable from the late-time attractor of the $\Lambda$CDM model. Therefore, the primary phenomenological signature of the model is not a persistent deviation from $\Lambda$CDM, but rather a transient phantom regime accompanied by a smooth phantom-divide crossing, while preserving the successful background evolution required by current cosmological observations.\\

{$\bullet$ \bf Model II:} We constrain the second model with the same three independent combinations of cosmological observation datasets. The free parameters sampled in the MCMC analysis are $({H_0,x_{10},x_{30},r_d,\alpha,\Lambda})$, with the adopted prior ranges listed in Table~\ref{tab:prior_range}. As discussed in the previous section, the dynamical variables are restricted to the physically admissible domain $x_1,x_3\geq0$. Moreover, the requirement that the quantity $\Gamma$ remains real imposes the conditions $\alpha\neq0$ and $\Lambda\leq0$, which are explicitly incorporated into the parameter priors. The exploration of the parameter space is therefore confined to regions yielding physically viable cosmological solutions. The resulting marginalized posterior distributions are presented in Fig.~\ref{fig3}, while the corresponding mean values and 68\% confidence intervals are summarized in Table~\ref{tab:best_fit_values}.

The three observational combinations also provide consistent constraints on the cosmological parameters. For the Pantheon Plus + CMB + DESI DR2 dataset, the model yields $H_0\simeq69.68$ together with a present matter density parameter $\Omega_{m}\simeq0.29$. The DES Y5 + CMB + DESI DR2 combination gives $H_0\simeq68.73$ and $\Omega_{m}\simeq0.30$, while the CMB + DESI DR2 dataset favors $H_0\simeq69.0$ with $\Omega_{m}\simeq0.30$. In all three analyses, the anisotropy parameter is more tightly constrained to $x_{30}\sim10^{-5}$ as compared to Model I, implying that any deviation from isotropy is extremely small and well within current observational limits.

\begin{figure}[t]
	\includegraphics[scale=0.5]{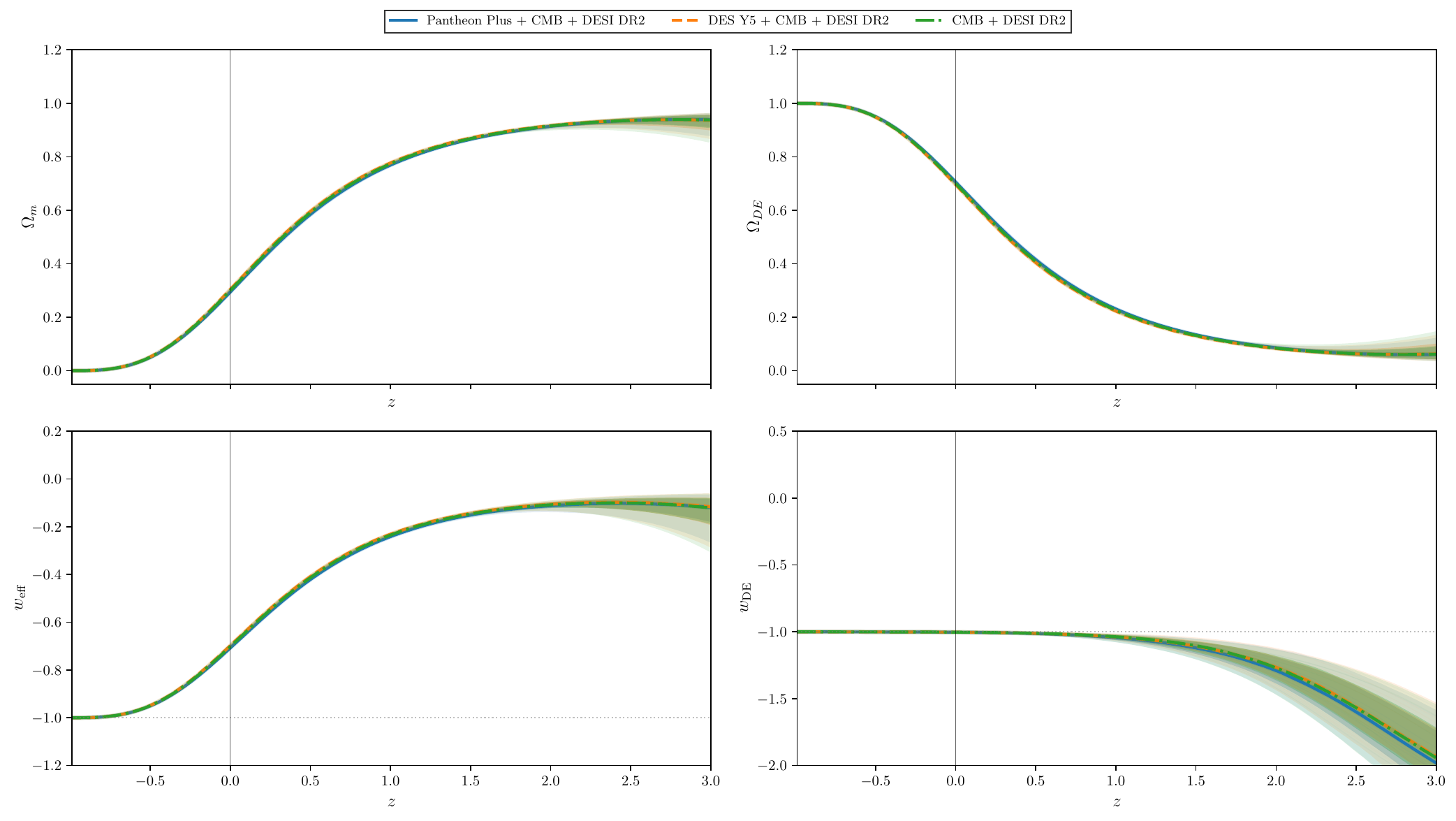}
	\caption{Evolution of the matter density parameter $\Omega_m$, dark energy density parameter $\Omega_{DE}$, effective equation of state parameter $w_{eff}$, and dark energy equation of state parameter $w_{DE}$ for Model II. The blue, orange, and green curves correspond to the Pantheon Plus+CMB+DESI DR2, DES Y5+CMB+DESI DR2, and CMB+DESI DR2 datasets, respectively. The associated dark and light shaded regions indicate the 68\% $(1\sigma)$ and 95\% $(2\sigma)$ confidence intervals.}
	\label{fig:params2}
\end{figure}

\begin{figure}[H]
    \centering
    \includegraphics[scale=0.5]{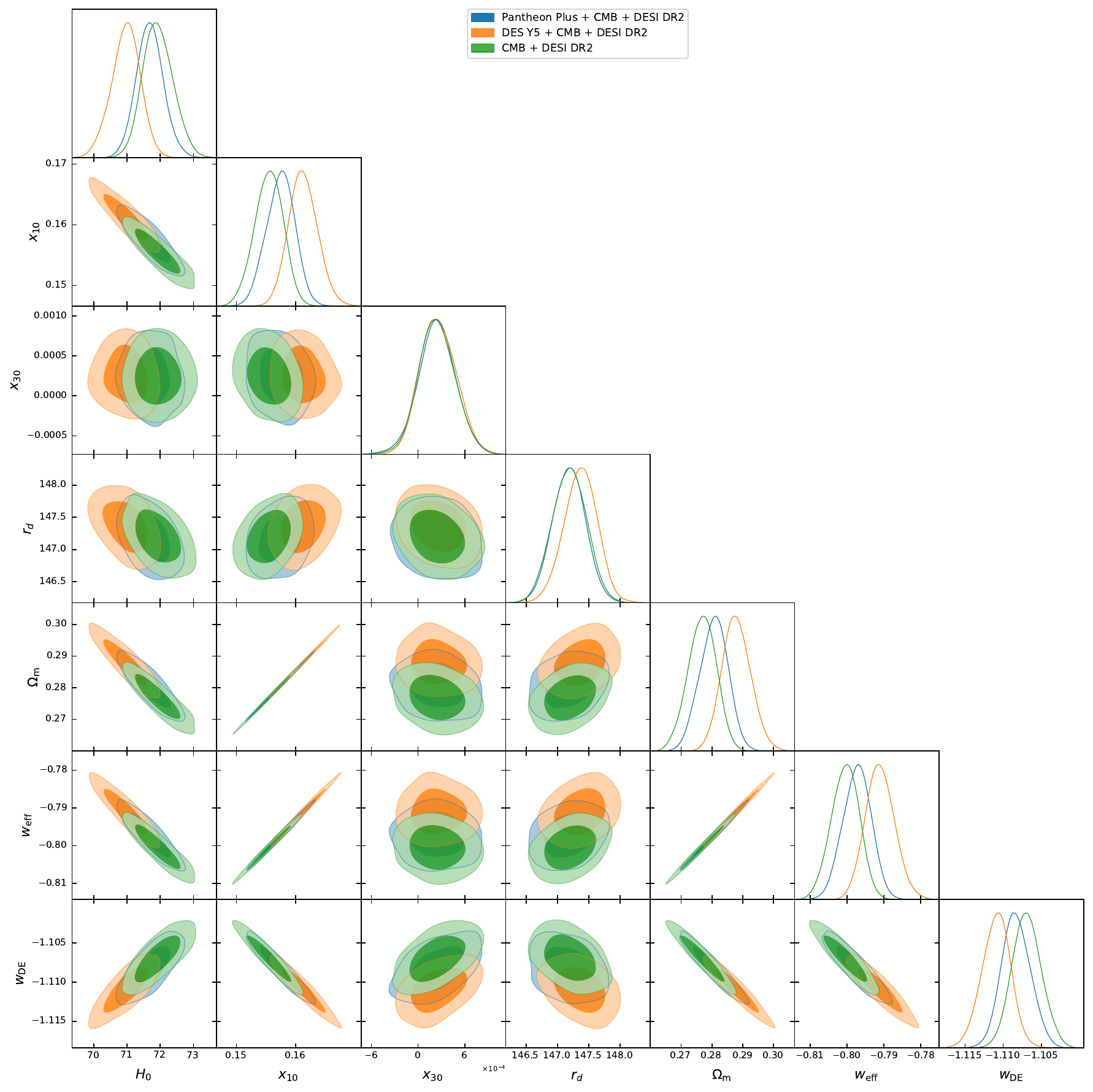}
    \caption{2D marginalized posterior distributions of the parameters of Model I for the Pantheon Plus+CMB+DESI DR2, DES Y5+CMB+DESI DR2, and CMB+DESI DR2 datasets. The contours indicate the 68\% and 95\% confidence levels.}
    \label{fig1}
\end{figure}

\begin{figure}[H]
    \centering
    \includegraphics[scale=0.5]{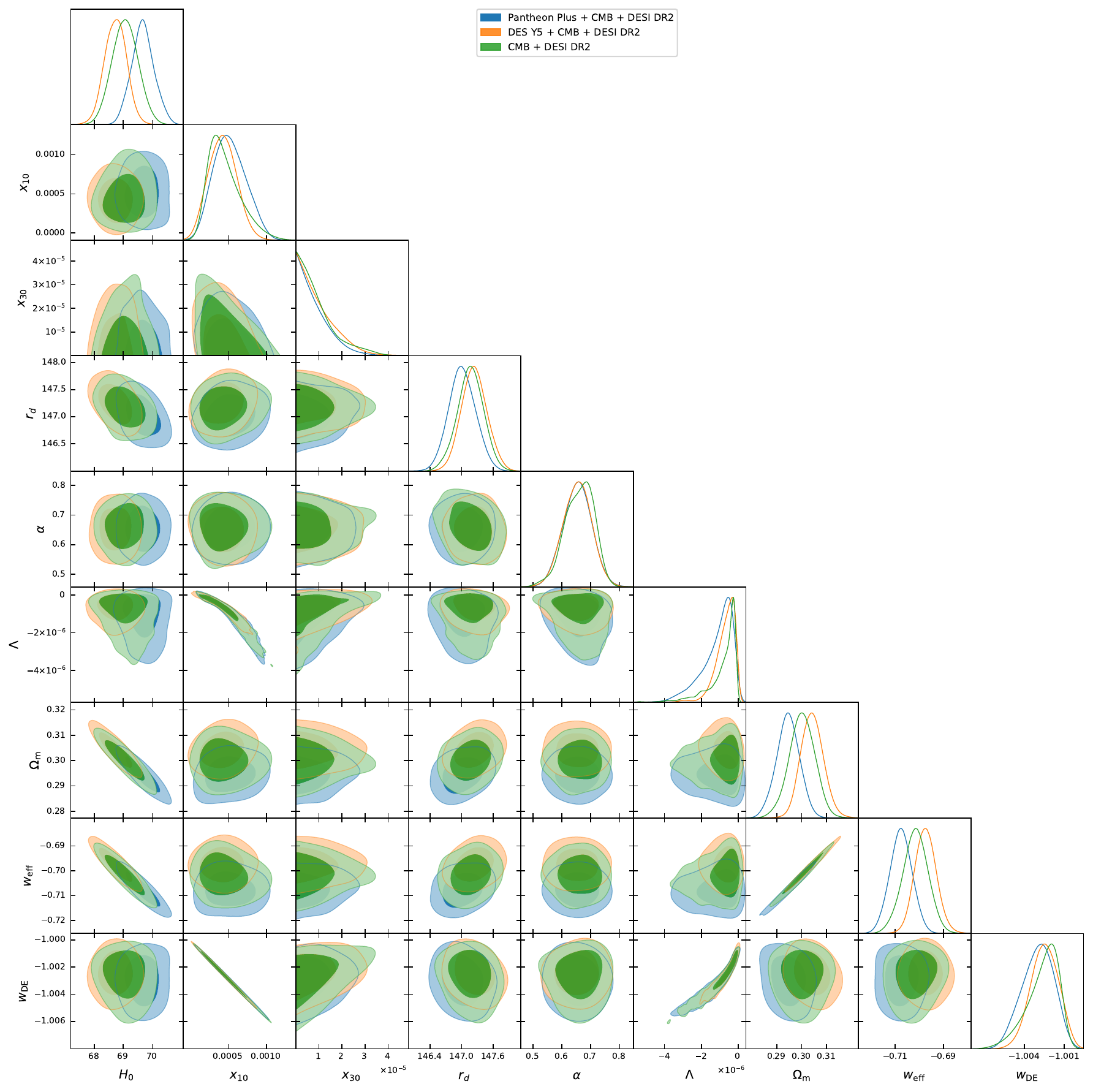}
    \caption{2D marginalized posterior distributions of the parameters of Model II for the Pantheon Plus+CMB+DESI DR2, DES Y5+CMB+DESI DR2, and CMB+DESI DR2 datasets. The contours indicate the 68\% and 95\% confidence levels.}
    \label{fig3}
\end{figure}

Another noteworthy outcome is that an effective EoS parameter is remarkably consistently constrained to $w_{ eff}\simeq-0.70$, confirming that the model successfully accounts for the present accelerated expansion of the universe. Likewise, the dark-energy EoS parameter is found to be $w_{DE}\simeq-1$ for all three dataset combinations, indicating that the dark-energy sector is observationally indistinguishable from a cosmological constant at the present epoch. The corresponding posterior distributions are shown in Fig.~\ref{fig3}, while their redshift evolution is illustrated in Fig.~\ref{fig:params2}.

The redshift evolution of $w_{DE}$ exhibits that at high redshift, $w_{ DE}$ resides in the phantom regime $w_{ DE}<-1$, indicating that the effective dark-energy component differs from a cosmological constant during the early stages of cosmic evolution. As the universe expands, $w_{ DE}$ evolves smoothly toward the cosmological constant boundary and reaches the phantom divide around $z\simeq0.6$. At the present epoch $z=0$, it converges to $w_{DE }\simeq-1$, corresponding to a de Sitter phase, and subsequently remains very close to this value throughout the future evolution. Consequently, unlike Model I, the phantom behavior in Model II is confined to earlier epochs, while the present and future universe are effectively described by a cosmological constant like a dark-energy component.

The evolution of $w_{eff}$ further supports this picture. During the matter-dominated era, $w_{ eff}\approx0$, reproducing the standard cosmological evolution expected for pressureless matter. As the contribution from dark energy gradually becomes dominant, $w_{ eff}$ decreases smoothly, reaching $w_{ eff}\simeq-0.70$ at the present epoch, in excellent agreement with the expectations of the $\Lambda$CDM model. The matter density parameter evolves monotonically from a matter-dominated early universe, with $\Omega_{m}\to1$, to the observed value
$\Omega_{m}\simeq0.30$ at the present epoch and vanishes asymptotically in the future. Conversely, the dark energy density parameter increases from a negligible early-time contribution to $\Omega_{DE}\simeq0.70$
today and approaches unity at late times, confirming the emergence of a dark-energy-dominated accelerated phase. Overall, Model II provides a cosmological evolution that remains remarkably close to the standard $\Lambda$CDM scenario despite the inclusion of a small anisotropic contribution $\sim10^{-5}$. The principal distinction from the standard model lies in the early-time phantom behavior of the effective dark-energy sector, which gradually relaxes to a de Sitter state. Consequently, the model preserves the successful late-time phenomenology of $\Lambda$CDM while allowing for a non-trivial dynamical evolution of dark energy at intermediate and high redshifts.

{$\bullet$ \bf Statistical Viability:}

To assess the statistical performance of the proposed models relative to the standard $\Lambda$CDM cosmology, we employ the minimum chi-square, $\chi^2_{\rm min}$, and the Akaike Information Criterion (AIC) \cite{aiccretria}. While $\chi^2_{\rm min}$ quantifies the goodness of fit, the AIC accounts for model complexity by penalizing the number of free parameters, thereby providing a balanced criterion for model comparison. The AIC is defined as ${AIC}=\chi^2_{ min}+2k$, where $k$ is the number of free parameters. The statistical results are summarized in Table~\ref{aic}. For Model~I, the obtained values $6<\Delta AIC<10$ indicate strong statistical evidence in favor of $\Lambda$CDM model over Model~I for all considered dataset combinations. On the other hand, Model~II yields $2< \Delta AIC<6$, corresponding to a weaker statistical preference for $\Lambda$CDM.
\begin{table}[t]
	\centering
	\begin{tabular}{l r | l r}
		\hline
		\multicolumn{2}{c}{\bf Model I} & \multicolumn{2}{c}{\bf Model II}\\
		\hline
		Parameters & Range & Parameters & Range\\
		\hline
		\(H_0\)  & $[60,80]$ & \(H_0\)  & $[60,80]$\\
		$x_{10}$  & $[0,1]$ & $x_{10}$  & $[0,1]$ \\
		$x_{30}$ & $[0,0.3]$ & $x_{30}$ & $[0,0.3]$ \\
		$r_d$ & $[130, 180]$ & $r_d$ & $[130, 180]$\\
            &               & $\alpha$ & $[-1, 1]$\\
            &               & $\Lambda$ & $[-2, 0]$\\

		\hline
		\hline
	\end{tabular}
	\caption{The prior range of the model parameters. }
	\label{tab:prior_range}
\end{table}

\begin{table}
	\centering
	\tiny
	\begin{tabular}{l c c c c c c c c c}
		\hline
		\multicolumn{8}{c}{\bf Datasets: Pantheon Plus + CMB + DESI DR2}\\
		\hline
		\hline
		Model & $H_0$ & $\Omega_{m}$ & $r_d$ & $x_{10}$ & $x_{30}$ & $\alpha$ & $\Lambda$ & $w_{\rm eff}$ & $w_{\rm DE}$\\
		\hline
		{\bf Model I:} 
		& $71.74 \pm 0.4$ 
		& $0.28 \pm 0.004$ 
		& $147.2 \pm 0.3$ 
		& $0.16 \pm 0.002$ 
		& $0.00023 \pm 0.00021$ 
        &
        &
		& $-0.79 \pm 0.004$ 
		& $-1.10 \pm 0.002$ \\
		
		{\bf Model II:} 
		&$69.68 \pm 0.4$  
		& $0.29 \pm 0.004$  
		& $147.0 \pm 0.2$  
		& $0.00052^{+00010}_{-00024}$ 
		&$(0.65 \pm 0.64)\times10^{-5}$
        &$0.65 \pm 0.05$
        &$(-0.90 \pm 0.85)$ $\times 10^{-6}$
		& $-0.70 \pm 0.004$ 
		& $-1.00 \pm 0.001$ \\
		
		{\boldmath\bf $\Lambda$CDM:} 
		&$69.50 \pm 0.4$
		& $0.29 \pm 0.004$ 
		& $146.9 \pm 0.2$ 
		& 
        &
        &
        &
       & $-0.70 \pm 0.004$ 
	    & $-1$ \\
		
		\hline
		\multicolumn{8}{c}{\bf Datasets: DES Y5 + CMB + DESI DR2}\\
		\hline
		\hline
		
		{\bf Model I:} 
		& $70.96 \pm 0.4$ 
		& $0.29 \pm 0.004$ 
		& $147.3 \pm 0.2$ 
		& $0.16 \pm 0.003$ 
		& $0.00026^{+0.0021}_{-0.0025}$
        &
        &
		& $-0.79 \pm 0.004$
		& $-1.11 \pm 0.002$\\
		
		{\bf Model II:} 
		&$68.73 \pm 0.4$  
		& $0.30 \pm 0.004$  
		& $147.1 \pm 0.2$  
		& $0.00042 \pm 0.00018$ 
		&$(0.76 \pm 0.78)\times10^{-5}$
        &$0.65 \pm 0.05$
        &$(-0.59 \pm 0.54)$ $\times 10^{-6}$
		& $-0.69 \pm 0.004$ 
		& $-1.00 \pm 0.001$ \\
		
		{\boldmath\bf $\Lambda$CDM:} 
		&$68.62 \pm 0.4$
		& $0.30 \pm 0.004$ 
		& $147.1 \pm 0.2$ 
		& 
        &
        &
        &
       & $-0.69 \pm 0.004$ 
	    & $-1$ \\
		
		\hline
		\multicolumn{8}{c}{\bf Datasets: CMB + DESI DR2}\\
		\hline
		\hline
		
		{\bf Model I:} 
		& $71.93 \pm 0.4$ 
		& $0.27 \pm 0.004$ 
		& $147.2^{+0.3}_{-0.2}$ 
		& $0.15 \pm 0.002$ 
		& $0.00024^{+0.00022}_{-0.00021}$
        &
        &
		& $-0.80 \pm 0.003$
		& $-1.10 \pm 0.002$\\
		
		{\bf Model II:} 
		&$69.06 \pm 0.4$  
		& $0.30 \pm 0.005$  
		& $147.2 \pm 0.2$  
		& $0.00045^{+0.00010}_{-0.00025}$ 
		&$(0.72 \pm 0.79)\times10^{-5}$
        &$0.66 \pm 0.05$
        &$(-0.56 \pm 0.85)$ $\times 10^{-6}$
		& $-0.70 \pm 0.005$ 
		& $-1.00 \pm 0.001$ \\
		
		{\boldmath\bf $\Lambda$CDM:} 
		&$69.00 \pm 0.4$
		& $0.29 \pm 0.004$ 
		& $147.0 \pm 0.2$ 
		& 
        &
        &
        &
       & $-0.70 \pm 0.004$ 
	    & $-1$ \\
		
		\hline
		\hline
	\end{tabular}
	\caption{Marginalized constraints on the cosmological model parameters obtained from the MCMC analysis for Model I, Model II, and the $\Lambda$CDM model. The results are reported as the median values with their corresponding 68\% confidence intervals for three observational dataset combinations: Pantheon Plus+CMB+DESI DR2, DES Y5+CMB+DESI DR2, and CMB+DESI DR2.}
	\label{tab:best_fit_values}
\end{table}

\begin{table}
	\centering
	\tiny
	\begin{tabular}{l c c c c c c}
		\hline
		\multicolumn{5}{c}{\bf Datasets: Pantheon Plus + CMB + DESI DR2}\\
		\hline
		\hline
		Model & $\chi_{ min}^{2}$ & $\Delta\chi_{min}^{2}=\chi_{min}^{2}-\chi_{ min, \Lambda CDM}^{2}$ & $AIC$  &$\Delta{AIC}={AIC}-{ AIC}_{\Lambda CDM}$  & \\
		\hline
		{\bf Model I:} 
		& $1485.61$ 
		& $6.90$ 
		& $1493.61$ 
		& $8.90$ 
         \\
		
		{\bf Model II:} 
		&$1477.69$  
		& $-1.02$  
		& $1489.69$  
		& $4.98$ 
        \\
		
		{\boldmath\bf $\Lambda$CDM:} 
		&$1478.71$
		& $0$ 
		& $1484.71$ 
		& $0$
        \\
		
		\hline
        \multicolumn{5}{c}{\bf  Datasets: DES Y5 + CMB + DESI DR2}\\
		\hline
        \hline
		{\bf Model I:} 
		& $1658.52$ 
		& $7.47$ 
		& $1666.52$ 
		& $9.47$ 
         \\
		
		{\bf Model II:} 
		&$1650.12$  
		& $-0.93$  
		& $1662.12$  
		& $5.07$ 
        \\
		
		{\boldmath\bf $\Lambda$CDM:} 
		&$1651.05$
		& $0$ 
		& $1657.05$ 
		& $0$
        \\
        \hline
         \multicolumn{5}{c}{\bf Datasets: CMB + DESI DR2}\\
		\hline
        \hline
		{\bf Model I:} 
		& $7.31$ 
		& $6.40$ 
		& $15.31$ 
		& $8.4$ 
         \\
		
		{\bf Model II:} 
		&$0.29$  
		& $-0.62$  
		& $12.29$  
		& $5.38$ 
        \\
		
		{\boldmath\bf $\Lambda$CDM:} 
		&$0.91$
		& $0$ 
		& $6.91$ 
		& $0$
        \\
        \hline
		\hline
	\end{tabular}
	\caption{Statistical comparison of Model I, Model II, and the $\Lambda$CDM model using the $\chi^2_{min}$,$\Delta\chi^2_{min}$, AIC, and $\Delta$AIC values obtained from MCMC analyses with the Pantheon Plus+CMB+DESI DR2, DES Y5+CMB+DESI DR2, and CMB+DESI DR2 datasets.}
	\label{aic}
\end{table}


\section{Conclusion}\label{VI}
In this work, we have investigated the cosmological dynamics of Bianchi-I spacetime in the framework of symmetric teleparallel $f(Q)$ gravity. By introducing suitable dimensionless variables, the modified field equations were recast into an autonomous dynamical system, allowing a systematic analysis of two viable $f(Q)$ models. We constrained both models using three independent combinations of cosmological observations: (i) Pantheon Plus + CMB + DESI DR2, (ii) DES Y5 + CMB + DESI DR2, and (iii) CMB + DESI DR2.

For both models, the MCMC analysis yields mutually consistent constraints across all dataset combinations, demonstrating the robustness of the obtained parameter estimates. The inferred values of the Hubble constant, matter density parameter, and sound horizon remain compatible with current observational bounds. Moreover, the anisotropic contribution is found to be tightly constrained, with $\sim10^{-4}$ for Model I and $\sim10^{-5}$ for Model II, indicating that any departure from isotropy is extremely small while remaining compatible with present cosmological observations.

The cosmological evolution of the two models exhibits distinct phenomenological signatures. Model I predicts a present accelerated universe with $w_{ eff}\simeq-0.79$ and a mildly phantom dark-energy EoS, $w_{\rm DE}\simeq-1.10$. The evolution of $w_{ DE}$ shows a smooth crossing of the phantom divide line, followed by a gradual convergence towards $w_{DE}=-1$ in the asymptotic future. Consequently, the phantom behavior is transient, and the late-time evolution naturally approaches a de Sitter phase that is observationally indistinguishable from the $\Lambda$CDM attractor.

In contrast, Model II remains considerably closer to the standard cosmological scenario. Although the dark-energy sector exhibits phantom behavior at earlier epochs, it smoothly evolves toward the cosmological constant boundary, reaching $w_{ DE}\simeq-1$ around the present epoch and remaining close to this value thereafter. Simultaneously, the effective EoS evolves from the matter-dominated regime to $w_{eff}\simeq-0.70$, while the matter density decreases monotonically to its observed present value before asymptotically vanishing in the far future. The resulting cosmic evolution closely reproduces the successful background expansion history of the $\Lambda$CDM model.

Overall, our analysis shows that anisotropic $f(Q)$ gravity remains a viable candidate for extending the standard cosmological framework. Both models recover the standard thermal history and matter-dominated epoch, while introducing characteristic modifications in the late-time dark-energy dynamics. Model I exhibits a transient phantom phase with a smooth phantom-divide crossing, whereas Model II displays an early phantom behavior that gradually approaches the cosmological-constant boundary. Combined with our previous qualitative critical-point analysis \cite{ghulam2025}, the present observational constraints support the existence of the theoretically predicted late-time de Sitter attractor. The MCMC results further indicate that both models reproduce the observed background expansion history while strongly suppressing anisotropic deviations. Despite their different transient evolutions, both models converge asymptotically toward a de Sitter state, demonstrating consistency between dynamical stability and observational viability. These characteristic dark-energy signatures may provide useful observational discriminants in upcoming precision cosmological surveys.


\section*{CONTOUR PLOTS}

\end{document}